\documentclass[a4paper,aps,pra,groupedaddress,twocolumn]{revtex4-1}      
\usepackage[utf8]{inputenc}  
\usepackage[T1]{fontenc}     
\usepackage[british]{babel}  
\usepackage[sc,osf]{mathpazo}
\usepackage[scaled=0.86]{berasans}  
\usepackage[colorlinks=true, citecolor=blue, urlcolor=blue]{hyperref}  
\usepackage{graphicx} 
\usepackage[babel]{microtype}  
\usepackage{amsmath,amssymb,amsthm,bm,amsfonts,mathrsfs,bbm} 
\usepackage{xspace}  
\usepackage{pgfplots}  

\def\be{\begin{equation}}
	\def\ee{\end{equation}}
\def\bea{\begin{eqnarray}}
	\def\eea{\end{eqnarray}}
\def\ben{\begin{equation*}}
	\def\een{\end{equation*}}
\def\bean{\begin{eqnarray*}}
	\def\eean{\end{eqnarray*}}
\def\bma{\begin{mathletters}}
	\def\ema{\end{mathletters}}
\def\bi{\begin{itemize}}
	\def\ei{\end{itemize}}

\DeclareMathOperator{\tr}{tr}

\begin{document}
\title{Optimal quantum violation of Clauser-Horne-Shimony-Holt like steering inequality}

\author{Arup Roy}
\email{arup145.roy@gmail.com}
\affiliation{Physics and Applied Mathematics Unit, Indian Statistical Institute, 203 B. T. Road, Kolkata 700108, India.}
\author{Some Sankar Bhattacharya}
\affiliation{Physics and Applied Mathematics Unit, Indian Statistical Institute, 203 B. T. Road, Kolkata 700108, India.}
\author{Amit Mukherjee}
\affiliation{Physics and Applied Mathematics Unit, Indian Statistical Institute, 203 B. T. Road, Kolkata 700108, India.}
\author{Manik Banik}
\affiliation{Physics and Applied Mathematics Unit, Indian Statistical Institute, 203 B. T. Road, Kolkata 700108, India.}

\begin{abstract}
We study a recently proposed Einstein-Podolsky-Rosen steering inequality [\href{http://arxiv.org/abs/1412.8178}{\it arXiv- 1412.8178 (2014)}]. Analogous to Clauser-Horne-Shimony-Holt (CHSH) inequality for Bell nonlocality, in the simplest scenario, i.e., 2 parties, 2 measurements per party and 2 outcomes per measurement, this newly proposed inequality has been proved to be necessary and sufficient for steering. In this article, using an equivalence between measurement incompatibility (non joint measurability) and steering, we find the optimal violation amount of this inequality in quantum theory. Interestingly, the optimal violation amount matches with optimal quantum violation of CHSH inequality, i.e., Cirel'son quantity. We further study the optimal violation of this inequality for different bipartite quantum states. To our surprise we find that optimal violation amount is different for different $2$-qubit pure entangled states, which is not the case for all other existing steering inequalities.     
      
\end{abstract}

\maketitle
\section{Introduction}
The phenomenal argument by Einstein, Podolsky and Rosen (EPR) in 1935 \cite{epr} to demonstrate the incompleteness of quantum mechanics, struck Schr\"{o}dinger with the concept of `steering' \cite{schr}. However, only recently, Wiseman \emph{et al.} have formalized the concept of steering in the form of a task \cite{wisemanprl07,wisemanpra07}. The task of steering can be seen as one's inability to construct a \emph{local hidden variable- local hidden state} (LHV-LHS) model that reproduces a given bipartite correlation. The work of Wiseman \emph{et al.} has generated an immense interest in the study of this steering phenomenon \cite{jevtic2014,rudolph2014,jevtic12014,brunneroneway2014,wittman2014,wiseman2014}. On the other hand, the concept of steering has been extended for multipartite case and the idea of n-partite genuine mutipartite steering have been explored \cite{reidg}. Unlike the two well studied non classical correlations, namely \emph{nonlocality} \cite{wehner2014} and \emph{entanglement} \cite{horodecki2009}, there is an inherent asymmetry in the task of steering. This is because in case of steering, on one subsystem (which is being `steered') the statistics must arise out of a valid measurement on a valid quantum state but no such constraint is required for the other subsystem. Here in addition to the simplicity of Bell's assumptions of \emph{local causality} one must also perform trusted measurements on one subsystem, whereas the other subsystem need not be trusted \cite{wiseman2013}. 

Apart form the foundational interest, the study of steering also finds applications in semi device independent scenario where only one party has trust on his/her quantum device but the other party's device is untrusted. As a concrete example it has been shown that steering allows for secure quantum key distribution (QKD) when one of the parties cannot trust their devices \cite{branciard2012}. One big advantage in this direction is that such scenarios are experimentally less demanding than fully device-independent protocols (where both of the parties distrust their devices) \cite{branciard2012} and, at the same time, require less assumptions than standard quantum cryptographic scenarios \cite{Bennett1992}.

In 1964, Bell sought a way to demonstrate that certain correlations appearing in quantum mechanics are incompatible with the notions of \emph{locality} and \emph{reality} aka \emph{local-realism} (LR), through an inequality involving measurement statistics \cite{Bell64}. A violation of such inequality implies the usefulness of correlations for EPR argument. In 1969, Clauser-Horne-Shimony-Holt (CHSH) proposed a set of simple Bell inequalities which are easy to realize experimentally \cite{CHSH69}. In the same spirit of Bell's inequality in non-locality, several steering inequalities (SI) have been proposed \cite{reid,ou,caval,wal,wal1}, so that a violation of any such steering inequalities can render a correlation to be steerable. But an unavoidable hindrance to formalize such SIs follows from the fact that steering scenario is device-independent only on one-side.

Recently Cavalcanti \emph{et al.} have proposed a CHSH-like inequality for quantum steering \cite{caval15}. They have derived an EPR-steering inequality that is necessary and sufficient for a set of correlations in the simplest two-party scenario involving two measurement settings per site and two outcomes per measurement, with mutually unbiased measurements at trusted party. In this article we derived the tight optimal quantum violation of the EPR-steering inequality proposed in \cite{caval15}. We have also studied the violation amount of this inequality for different well-known classes of $2$-qubit states. Interestingly, we find that for different $2$-qubit pure entangled states the optimal violation amount of this inequality differ. Note that, for all other existing SIs such difference is not observed.

The organization of this article goes as follows: we first briefly review few existing steering inequalities along with the newly proposed CHSH like steering inequality. Then we show that the optimal violation of this CHSH like steering inequality in quantum theory is restricted to $2\sqrt{2}$. Then we study how the optimal violation amount of the CHSH like inequality varies for different entangled states.

\section{Steering Inequalities}
To test EPR steering Reid first proposed a testable formulation for continuous-variable systems based on position-momentum uncertainty relation \cite{reid} which was experimentally tested by Ou et al \cite{ou}. Cavalcanti \emph{et al.} developed a general construction of experimental EPR-steering criteria based on the assumption of existence of LHS model \cite{caval}. Importantly this general construction is applicable to both the discrete as well as the continuous-variable observables and Reid’s criterion appears as a special case this general formulation. On the other hand, Walborn \emph{et al.} formulated a steering inequality based on Bialynicki-Birula and Mycielski’s entropic position-momentum uncertainty relation \cite{wal,wal1}. As the entropic uncertainty relation implies Heisenberg’s uncertainty relation, hence the set of states violating Walborn \emph{et al.}’s steering inequality contains all the states violating Reid’s inequality. Thus Walborn \emph{et al.}’s steering criterion is more powerful than Reid's one. However this is true for continuous variable case only, not for the discrete case. In all instances the violation of the aforementioned inequalities by different pure entangled states are the same.

All of these steering inequalities have been proved to be as sufficient conditions for witnessing steering in bipartite quantum systems. But none of these condition is supposed to be a necessary and sufficient condition for steering. Search for such an necessary and sufficient condition has been culminated in a recent development by Cavalcanti \emph{et al.} \cite{caval15}. They have proposed an CHSH like EPR-steering inequality that is necessary and sufficient for the set of correlations in the simplest scenario involving two settings and two outcomes per setting, with mutually unbiased measurements at trusted end. At this point it is interesting to ask the following questions: 
\begin{itemize}
\item[(a)] what is the optimal violation of this newly proposed CHSH like steering inequality in quantum theory?
\item[(b)] how the violation amount of the concerned inequality depends on the state?
\end{itemize}
We provide a definite answer for the first question and study the second one for some classes of states.

\section{Maximum violation of CHSH-like steering inequality}
Let us first briefly review the steering scenario as introduced by Wiseman \emph{et al.} \cite{wisemanprl07,wisemanpra07}. Given a pair of systems at Alice and Bob, denote $\mathcal{D}_{\alpha}$ and $\mathcal{D}_{\beta}$ the sets of observables in the Hilbert space of Alice’s and Bob’s system, respectively. An element of $\mathcal{D}_{\alpha}$ is denoted by $A$, with a set of outcomes labeled by $a\in\mathcal{L}(A)$, and similarly for Bob. The joint state $\rho_{AB}$ of the system is steerable by Alice iff it is not the case that for all $a\in\mathcal{L}(A)$, $b\in\mathcal{L}(B)$, $A\in\mathcal{D}_{\alpha}$, $B\in\mathcal{D}_{\beta}$, the joint probability distributions can be written in the form
\begin{equation}\label{steer}
P(a,b|A,B;\rho_{AB})=\sum_{\lambda}\wp(\lambda)\wp(a|A,\lambda)P(b|B;\rho_{\lambda}),
\end{equation}
where $\wp(a|A,\lambda)$ denotes an arbitrary probability distribution and $P(b|B;\rho_{\lambda})$ denotes the quantum probability of outcome $b$ given measurement $B$ on state $\rho_{\lambda}$. In other words the state $\rho_{AB}$ will be called steerable if it does not satisfy a LHV-LHS model. Note that, if for a given measurement strategy the correlations have a LHV–LHS model, this does not imply that the underlying state is not steerable, since there could be another strategy that does not. In the simplest scenario where Alice and Bob each have a choice between two dichotomic measurements to perform: $\{A_1,A_2\}$, $\{B_1,B_2\}$, and outcomes of $A$ are labeled $a\in\{-1,+1\}$ and similarly for the other measurements, the authors of ref.\cite{caval15} derived a necessary and sufficient criterion for steering which reads as:
\begin{multline}
	S=\sqrt{\left\langle (A_{1}+A_{2})B_{1}\right\rangle ^{2}+\left\langle (A_{1}+A_{2})B_{2}\right\rangle ^{2}}\\
	+\sqrt{\left\langle (A_{1}-A_{2})B_{1}\right\rangle ^{2}+\left\langle (A_{1}-A_{2})B_{2}\right\rangle ^{2}} \leq2.\label{eq_nec_suf}
\end{multline}
We know that, in the simplest Bell scenario which involves two observers with $2$ dichotomic measurements per site, the set of local correlations lies in a polytope (LHV polytope) with CHSH inequality providing the non trivial facets of the LHV polytope. In the steering scenario with similar settings the set of correlations having LHV-LHS description also form a polytope.

One of the authors of this article, along with other collaborators, have shown that measurement incompatibility limits the Bell-CHSH inequality violation in quantum theory to Cirel'son bound \cite{manik}. Adopting similar approach and using a recently established connection between measurement incompatibility and steering,  we derive the optimal quantum violation of the steering inequality (\ref{eq_nec_suf}). Before establishing this result we first briefly review the concept of measurement incompatibility and the concept of unsharp measurement in quantum theory.

{\it Measurement incompatibility:}  In the case of projective measurements, compatibility is uniquely
captured by the notion of commutativity. Non commutative projective measurements in quantum mechanics do not admit unambiguous joint measurement \cite{Varadarajan'85}. With the introduction of the generalized measurement i.e. positive operator-valued measures (POVMs) \cite{Kraus'83,Nielsen'10}, it was shown that observables which do not admit perfect joint measurement, may allow joint measurement if the measurements are made sufficiently fuzzy \cite{Busch'85,Busch'86}. Therefore, for general measurements there is no unique notion of compatibility. In this article measurement incompatibility is captured by non joint measurability \cite{Busch'96}. 

Mathematically, a POVM consists of a collection of operators $\{A_{a|x}\}_a$ which are positive, $A_{a|x}\ge 0~\forall~a$, and sum up to the identity, $\sum_aA_{a|x}=\mathbf{1}$. Here $a$ denotes measurement outcome and $x$ denotes measurement choices. Physically, any POVM can be realized by first letting the physical system interact with an auxiliary system and then measuring an ordinary observable on the auxiliary system.  A set of $m$ POVMs $\{A_{a|x}\}_a$ is called jointly measurable if there exists a measurement $\{A_{\vec{a}}\}$ with outcome $\vec{a}=[a_{x=1},a_{x=2}, . . . ,a_{x=m}]$ where $a_x\in\{0,1,..,n\}$ gives the outcome of $x^{th}$ measurement, i.e.,
\begin{equation}
A_{\vec{a}}\ge 0,~~\sum_{\vec{a}}A_{\vec{a}}=\mathbf{1},~~\sum_{\vec{a}\backslash a_x}A_{\vec{a}}=A_{a|x}~\forall~x.\label{joint-meas}
\end{equation}
where $\vec{a}\backslash a_x$ stands for the elements of $\vec{a}$ except for $a_x$. Hence, all POVM elements $A_{a|x}$ are recovered as marginals of the mother observable $A_{\vec{a}}$. 

{\it Unsharp measurement:} Let us consider two dichotomic quantum measurements $A_1$ and $A_2$, which are not jointly measurable. Denoting eigenvalues of these operator as $\pm 1$, the expectation value over some state $\sigma$ can be expressed as:
\begin{equation*}
\langle A_k\rangle_{\sigma}= p(+1|A_k)-p(-1|A_k),~~k\in\{1,2\};
\end{equation*}
where $p(\pm 1|A_k)=\mbox{Tr}(A^{\pm}_k\sigma)$, with $A^{\pm}_k$ being the POVM elements corresponding to $\pm 1$. The unsharp of fuzzy observable is given by $A^{(\eta)}_k:=\{A^{\pm(\eta)}_k~|~A^{\pm(\eta)}_k\ge 0~\&~A^{+(\eta)}_k+A^{-(\eta)}_k=\mathbf{1}\}$, with
\begin{equation*}
A^{\pm(\eta)}_k=\frac{1\pm \eta}{2}A^{+}_k+\frac{1\mp\eta}{2}A^{-}_k.
\end{equation*}
Here $\eta\in(0,1]$ is known as ``unsharpness parameter" and the fraction $\frac{1+\eta}{2}$ is called ``degree of reality" \cite{Busch'86}. It may happen that the observables $A_1$ and $A_2$ do not allow any joint measurement, but with introduction of sufficient amount of unsharpness, there unsharp versions $A^{(\eta)}_1$ and $A^{(\eta)}_2$ may allow joint measurement. In ref.\cite{manik}, the authors have proved that 	given any d-dimensional quantum system, joint measurement for unsharp versions of any two dichotomous observables ${A_1}$ and ${A_2}$ of the system is possible with the largest allowed value of the unsharpness parameter $\eta_{opt}=\frac{1}{\sqrt{2}}$. Note that the expectation value of an unsharp observable $A^{(\eta)}_k$ over some quantum state $\sigma$ is related to the expectation value of its sharp version in the following manner,  
\begin{equation}
\langle A^{(\eta)}_k\rangle_{\sigma}=\eta\langle A_k\rangle_{\sigma}.
\end{equation}
Similarly, if Alice performs unsharp measurement $A^{(\eta)}_k$ on her part and Bob performs sharp measurement $B_j$ on his part of a bipartite shared state $\rho_{AB}$ than we have,
\begin{equation}\label{eta}
\langle A^{(\eta)}_kB_j\rangle_{\rho_{AB}}=\eta\langle A_kB_j\rangle_{\rho_{AB}}.
\end{equation} 

Except from quantum entanglement, another necessary ingredient which is necessary for study of quantum nonlocality is the existence of incompatible set of measurements. In the simplest bipartite scenario Wolf \emph{et al.} showed that any set of two incompatible POVMs with binary outcomes can always lead to violation of the CHSH-Bell inequality \cite{Wolf2009}. But, recently in refs.\cite{bru,ghu} the authors have proved that this result does not hold in the general scenario where numbers of POVMs and outcomes are arbitrary. However in this general settings the authors of \cite{bru,ghu} have established a connection between measurement incompatibility and a weaker form of quantum nonlocality i.e., EPR-Schr\"{o}dinger steering. They have shown that for any set of  incompatible POVMs (i.e. not jointly measurable), one can find an entangled state, such that the resulting statistics violates a steering inequality. Please note that, one of the authors of this article has recently proved that connection between measurement incomparability and steering holds for a more general class of tensor product theories rather than just Hilbert space quantum theory \cite{Manik'2015}. 

Let Alice performs a measurement assemblage $\{A_{a|x}\}$ on her part of a bipartite shared quantum state $\rho_{AB}$. Upon performing measurement $x$, and obtaining outcome $a$, the (un-normalized) state held by Bob is given by $\sigma_{a|x}=\mbox{Tr}(A_{a|x}\otimes\mathbf{1}\rho_{AB})$. The normalized state on Bob side is given by $\sigma_{a|x}/\mbox{Tr}(\sigma_{a|x})$. Also we have $\sum_a\sigma_{a|x}=\sum_a\sigma_{a|x'}~\mbox{for}~x\ne x'$, which actually ensure no signaling from Alics to Bob. The state assemblage $\{\sigma_{a|x}\}$ is un-steerable iff it admits a decomposition of the form
\begin{equation}
\sigma_{a|x}=\pi(\lambda)p(a|x,\lambda)\sigma_{\lambda},~~\forall~a,x,
\end{equation} 
where $\sum_{\lambda}\pi(\lambda)=1$. Existence of such decomposition for state assemblage on Bob's side ensures that the statistics obtained from the state $\rho_{AB}$ admits a combined LHV-LHS model of the form of Eq.(\ref{steer}). The authors in refs.\cite{bru,ghu} have shown that the assemblage $\{\sigma_{a|x}\}$, with $\sigma_{a|x}=\mbox{Tr}(A_{a|x}\otimes\mathbf{1}\rho_{AB})$, is un-steerable for any state $\rho_{AB}$ acting on $\mathbb{C}^d\otimes\mathbb{C}^d$ if and only if the set of POVMs $\{A_{a|x}\}$ acting on $\mathbb{C}^d$ is jointly measurable. As a corollary of this result we can say that

{\it Corollary 1}: The assemblage $\{ \sigma_{a|x} \}$, with $\sigma_{a|x} = \tr_A(A_{a|x} \otimes \mathbf{1} \rho_{AB} )$ and $x\in \{1,2\}$, is unsteerable for any state $\rho_{AB}$ acting in $ \mathbb{C}^d \otimes \mathbb{C}^d$ if and only if the set of POVMs $\{ A_{a|x} \}$ acting on $\mathbb{C}^d$ is jointly measurable.

At this stage, we are now in a position to prove our main result, which is described in the following theorem.

{\it Theorem 2}: Consider a composite quantum system composed of two subsystem with state spaces $\mathcal{H}_1$ and $\mathcal{H}_2$, respectively. For any pair of dichotomic observables $A_1$, $A_2$ for the first system and the mutually unbiased dichotomic observables $B_1$, $B_2$ for the second system and the joint state $\rho_{AB}$ acting on $\mathcal{H}_1\otimes\mathcal{H}_2$, we have the following inequality:
\begin{equation}
		S \le \frac{2}{\eta_{opt}},
\end{equation}
where $\eta_{opt}$ is the optimal unsharpness parameter that allows joint measurement for any two dichotomic quantum observables.

{\it Proof}: Let us consider two arbitrary dichotomic observables $\{A_{a|x}\}$ on Alice side, $x\in\{1,2\}$ and $a\in\{-1,+1\}$. These two observables in general may not allow joint measurement. However, introduction of unsharpness makes it possible to measure the unsharp version of these two observables jointly. Let the optimal unsharpness is $\eta_{opt}$ which allows joint measurement for any two dichotomic observables.  

Now according to Corollary $1$, as far as observables on Alice's side are jointly measurable, they will not violate any steering inequality and hence the steering inequality (\ref{eq_nec_suf}). Thus we have
\begin{widetext}
\begin{eqnarray*}
\sqrt{\left\langle (A_{1}^{(\eta_{opt})}+A_{2}^{(\eta_{opt})})B_{1}\right\rangle ^{2}+\left\langle (A_{1}^{(\eta_{opt})}+A_{2}^{(\eta_{opt})})B_{2}\right\rangle ^{2}}~~~~~~~~~~~~~~~~~~~~~~~~~~~~~~~~~~~~~~~~~~~~\\
		+\sqrt{\left\langle (A_{1}^{(\eta_{opt})}-A_{2}^{(\eta_{opt})})B_{1}\right\rangle ^{2}+\left\langle (A_{1}^{(\eta_{opt})}-A_{2}^{(\eta_{opt})})B_{2}\right\rangle ^{2}} \leq 2.
\end{eqnarray*}
Now using the expressing from Eq.(\ref{eta}) we get,

\begin{eqnarray*}
\sqrt{\left\langle (A_{1}+A_{2})B_{1}\right\rangle ^{2}+\left\langle (A_{1}+A_{2})B_{2}\right\rangle ^{2}}~~~~~~~~~~~~~~~~~~~~~~~~~~~~~~~~~~~~~~~~~~\\
		+\sqrt{\left\langle (A_{1}-A_{2})B_{1}\right\rangle ^{2}+\left\langle (A_{1}-A_{2})B_{2}\right\rangle ^{2}} \leq\frac{2}{\eta_{opt}}.
\end{eqnarray*}
\end{widetext}
The value of $\eta_{opt}$ in quantum theory is proved to be $1/\sqrt{2}$ \cite{manik}. Therefore the upper bound of the steering inequality (\ref{eq_nec_suf}) in quantum theory is $2\sqrt{2}$, i.e., $S\le 2\sqrt{2}$, which is numerically equal to the celebrated Cirel'son value. Naturally the question arises whether this value is tight or not. In the following section we answer this question affirmatively by showing that sharing a maximally two qubit entangled state with suitable choice of measurement, this value can be achieved in quantum theory.    

\section{Violation of steering inequality by different states} 
In this section we study the optimal violation of the steering inequality (\ref{eq_nec_suf}) for different given entangled states.

{\it Observation 1}: $S^{opt}$ is different for different $2$-qubit pure entangled states. 

\begin{figure}[h!]
	\centering
	\includegraphics[height=4.5cm,width=7cm]{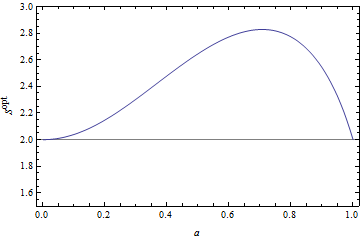}
	\caption{(Color on-line) The optimal violation of the steering inequality (\ref{eq_nec_suf}) for different $2$-qubit pure entangled states.}\label{fig1}
\end{figure}

Consider that an arbitrary 2-qubit pure entangled state $|\psi\rangle_{AB}=a|00\rangle_{AB}+b|11\rangle_{AB}~~(|a|^2+|b|^2=1)$ is shared between Alice and Bob. Alice performs measurements $A_{1}=\frac{1}{2}(\mathbb{I}+\vec{m}.\vec{\sigma})$ and $A_{2}=\frac{1}{2}(\mathbb{I}+\vec{n}.\vec{\sigma})$ on her part of the entangled particle where $|\vec{m}|,|\vec{n}|\le 1$. Similarly Bob performs measurements $B_{1}=\frac{1}{2}(\mathbb{I}+\vec{c}.\vec{\sigma})$ and $B_{2}=\frac{1}{2}(\mathbb{I}+\vec{d}.\vec{\sigma})$ on his part of the entangled particle where $|\vec{c}|,|\vec{d}|\le 1$ and $\vec{c}.\vec{d}=0$, i.e. $B_{1},B_{2}$ are mutually unbiased qubit measurements. Varying over the measurement directions of Alice and Bob we have numerically found the optimal violation amount  of the steering inequality (\ref{eq_nec_suf}) for a given $2$-qubit pure entangled state and in Fig.\ref{fig1} we have plotted it with respect to state parameter $a$. 
From Fig.\ref{fig1} it is clear that all $2$-qubit pure entangled states violate the steering inequality (\ref{eq_nec_suf}). It is important to note that the optimal violation amount is different for different states with maximally entangled state providing the maximum violation $2\sqrt{2}$ where Alice performs measurement in the directions $\vec{A_{1}}\equiv(-0.158719,-0.9556,-0.248268)$ and $\vec{A_{2}}\equiv(0.632668,0.0946074,-0.768622)$ and Bob's MUB measurements look like $\vec{B_{1}}\equiv(-0.268126,-0.697672,0.666859)$ and $\vec{B_{2}}\equiv(-0.597419,-0.659831,-0.455756)$. Please note that except from the steering inequality considered here, for all other existing steering inequalities the violation amount does not change with respect to different pure entangled states. 

{\it Observation 2}: $S^{opt}$ for $2$-qubit Werner class of states.  

\begin{figure}[h!]
	\centering
	\includegraphics[scale=0.35]{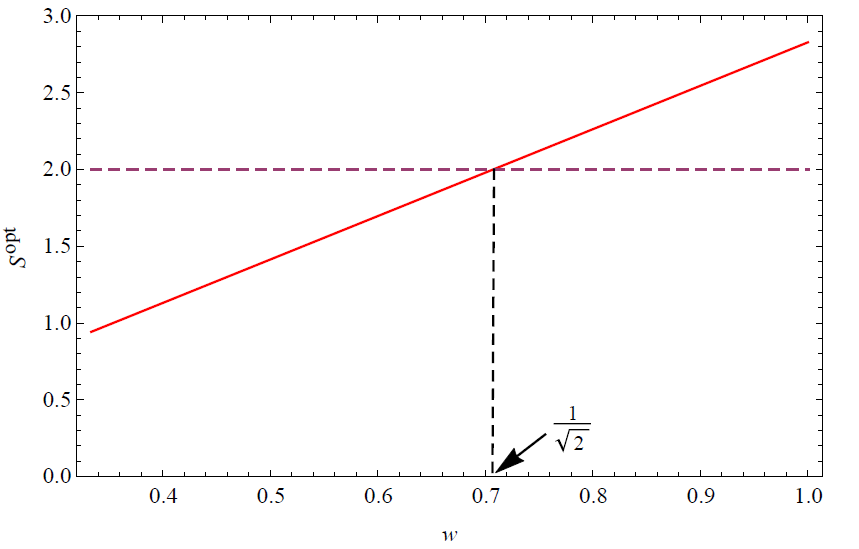}
	\caption{(Color on-line) The optimal violation of the steering inequality (\ref{eq_nec_suf}) for $2$-qubit Werner class of states.}\label{fig2}
\end{figure}

Consider that Alice and Bob share the Werner state $W^w_{AB} = w|\psi^-\rangle_{AB}\langle \psi^-|+(1-w)\frac{\mathbb{I}}{2}\otimes\frac{\mathbb{I}}{2}$; where $|\psi^-\rangle_{AB}=\frac{1}{\sqrt{2}}(|01\rangle_{AB}-|10\rangle_{AB})$ is the singlet state. Likewise in the previous case, varying over the measurement setup of Alice and Bob we find the optimal violation amount of the steering inequality (\ref{eq_nec_suf}) for different Werner state and plot this values in Fig.\ref{fig2}.
From Fig.\ref{fig2} it is clear that Werner states violate the steering inequality for $w>\frac{1}{\sqrt{2}}$ and the violation amount increase with the parameter $w$.
\section{Conclusion}
There have been several attempts to quantify quantum steering. Two most recent instances are `steerable weight' \cite{sky}, `steering robustness' \cite{pia} and `relative entropy of steering' \cite{gal}. But all of these proposed quantifiers assign same value for all pure entangled states \cite{gal}. At this point our observation becomes interesting. Since the amount of violation can also be shown to be non increasing under steering non-increasing operations (SNIOs) \cite{gal}, one can take this amount of violation to be a valid quantifier of quantum steering. Our study stipulates further studies whether some semi device independent protocol(s) can be deigned whose payoff scales with violation amount of the steering inequality considered here. In such case different violation amount of the steering inequality by different pure entangles states will have an operational explanation. 

We also find the violation of the CHSH like steering inequality by $2$-qubit Werner states. The inequality is violated by Werner states if $w>\frac{1}{\sqrt{2}}$ and beyond this value it follows a trait- the more entangled the state, more is the violation. We also find that the violation of this steering inequality in QM is tightly upper bounded by $2\sqrt{2}$, the well known Cirel'son quantity.

\textbf{Acknowledgment}:  
It is a pleasure to thank Prof. Guruprasad Kar for various simulating discussions and useful suggestions. We also thank Dr. Ashutosh Rai for useful discussions. AM acknowledge support from the CSIR project 09/093(0148)/2012-EMR-I. n. MB thankfully acknowledges discussions with Prof. Sibasish Ghosh during recent visit to the Institute of Mathematical Sciences, Chennai, India.

\end{document}